\documentclass[prl,twocolumn]{revtex4}

\usepackage{graphicx}

\begin{document}

\title{Cyclic Lattice Feshbach Resonances}
\author{Fei Zhou}
\affiliation{ \textit{PITP and Department of Physics and Astronomy,
 The University of British Columbia, \\
6224 Agricultural Road, Vancouver, B.C., Canada V6T1Z1}}
\date{{\small \today}}

\begin{abstract}
In this Letter we illustrate the possible
cyclic fermion pairing states across Feshbach resonances in optical 
lattices.
In cyclic fermion pairing, the pairing amplitude exhibits an oscillatory 
behavior as the detuning varies. We estimate the quasi-particle gaps in 
different regimes of the resonances.
\end{abstract}

\maketitle

Feshbach resonances in optical lattices were recently investigated in 
experiments \cite{Kohl05,Stoferle06}, after the earlier observation of 
Feshbach resonances in 
traps\cite{Greiner03,Jochim03,Zwierlein03,Strecker03}.
The fascinating phenomenon of lattice Feshbach resonances ({\em LFR}s) has 
attracted 
a lot of attention; both 
many-body physics and few-body aspects across {\em LFR}s 
have been studied\cite{Busch98,Carr05,Zhou05a,Diener05,Koetsier06}. 
Across {\em LFR}s 
fermionic atoms were observed to form shallow molecules, and 
superfluids should be 
present in optical lattices when tunnelling is permitted  
\cite{Stoferle06}. However, superfluids in {\em LFR}s can 
exhibit distinct properties. For instance, repulsive interactions between 
molecules\cite{Petrov05}, which have usually been overlooked in
standard theoretical studies of superfluids near Feshbach resonances 
in traps, play a significant role in {\em LFR}s; the suprefluid 
long-range-order can be 
substantially renormalized by these interactions\cite{Zhou05b}.

In this Letter, we study the pairing 
between fermions near {\em LFR}s, taking into account repulsive 
interactions between 
bosonic molecules in the closed channel. 
We show that in the limit of large repulsive interactions between 
molecules, or small molecule band width, {\em LFR}s should
exhibit a cyclic behavior as the detuning varies. 
Namely, the fermion quasi-particle energy 
gaps and critical temperatures of 
superfluids across {\em LFR}s exhibit an oscillatory behavior;
the gap reaches maxima in a {\em hard-core} boson limit.

Consider the following Fermi-Bose lattice model

\begin{eqnarray}
H&=& H_f + H_b + H_{bf}; \nonumber \\
H_f &=&-\sum_{<kl>, \eta,\eta',\sigma} t_{f, \eta\eta'}
( f^\dagger_{k\eta \sigma} f_{l\eta'\sigma} + h.c.)\nonumber \\
&+& \sum_{k, \eta, \sigma} (\epsilon_\eta -\mu) f^\dagger_{k\eta\sigma} 
f_{k\eta\sigma}
\nonumber \\
H_b&=&-t_b \sum_{<kl>} (b^\dagger_k b_l +h.c.)
\nonumber \\
&+&\sum_k (2v-2\mu)b^\dagger_kb_k
+U_b\sum_{k} \hat{n}_{bk} (\hat{n}_{bk}-1);
\nonumber \\
H_{bf}&=&-\gamma_0 \sum_{k,\eta} g_{\eta\eta'}
(b^\dagger_{k} f_{k\eta\uparrow}f_{k\eta'\downarrow}+h.c.)
\label{FBHM}
\end{eqnarray}
Here $k$, $\eta$ and $\sigma$ label lattice sites, on-site orbitals and
spins; $\eta=1,2,...M$, $\sigma=\uparrow,\downarrow$.
$f^\dagger_{k\eta\sigma}$ $(f_{k\eta\sigma})$ is the creation(annihilation) 
operator
of an atom (or fermion) at site $k$, with on-site orbital energy 
$\epsilon_\eta$ and spin 
$\sigma$.
$b^\dagger_k$($b_k$) is the creation 
(annihilation) operator of a molecule (or boson) at site $k$ (For 
simplicity, We assume there is 
only one bosonic orbital degree of freedom at each site).
The boson number operator is
$\hat{n}_{bk}=b^\dagger_kb_k$.
$t_{f,\eta\eta'}$ and $t_b$ are hopping integrals of fermions and bosons respectively and hopping
occurs over neighboring sites labeled as $<kl>$.
The single particle tunneling of bosons in optical lattices is 
assumed to be small, so that 
$t_b$ is much less than $U_b$,
the on-site repulsive interaction between bosonic molecules;
$t_f$ is much larger than the energy difference between two 
orbitals.
$H_b$ is the Hamiltonian previously introduced for atoms in optical lattices 
\cite{Jaksch98};
it has been used to study Mott insulator-superfluid phase 
transitions which were recently observed in optical 
lattices\cite{Greiner02}.

Finally
in narrow resonances, or in the large-N limit treated 
before\cite{Zhou05a,Zhou05b}, 
the matrix $g_{\eta\eta'}$ is proportional to
$\delta_{\eta\eta'}$.
However, in other limits the $g$-matrix generally has off-diagonal 
terms\cite{Diener05}. Here we assume $g_{\eta\eta'}$ to be an 
arbitrary symmetric matrix so that the spatial wavefunction of pairs 
is symmetric and spin wavefunction antisymmetric.

The self-consistent chemical potential is determined by the following 
formula,

\begin{eqnarray}
&& N=2 N_b(2\mu-2v; t_b, U_b; \lambda_b(\gamma_0, t_f, \mu)) \nonumber 
\\
&& +N_f(\mu; t_f;\lambda_f(\gamma_0, t_b, U_b))
\label{cpotential}
\end{eqnarray}
where $N$, $N_{b}$ and $N_f$ are, respectively, the total number of 
fermions,
the number of molecules or bound states of atoms and the number of atoms 
per lattice site.
$\lambda_{f(b)}$ is the effective interaction constant between fermionic 
atoms(molecules) mediated by molecules (fermions). 
We are interested in the limit where
the coupling strength $\gamma_0$ is small compared with $U_b$ 
so that in the calculation of 
$N_b (N_f)$ we can neglect the $\lambda_b (\lambda_f)$-dependence.

When $t_b$ is much smaller than $U_b$, 
molecules can be considered to be in a 
Mott insulating ground state
to the zeroth order 
in $\gamma_0$ \cite{remark2}.
Therefore $N_b$ is a step-like function of 
the detuning 
$v $. In the limit of zero $t_b$, 
$N_b$ is equal to an integer $n$ 
whenever $ |2\mu -2v/U_b - I| < 1/2$; 
$N_b$ has two degenerate solutions when $2\mu-2v/U_b=I + 1/2$
and the value of $N_b$ jumps from $I$ to $I+1$ at that particular detuning 
signifying a discontinuous transition
between two plateaus in the $N_b$ versus $v$ curve.

\begin{figure}[tbp]
\begin{center}
\includegraphics[width=3.3in]
{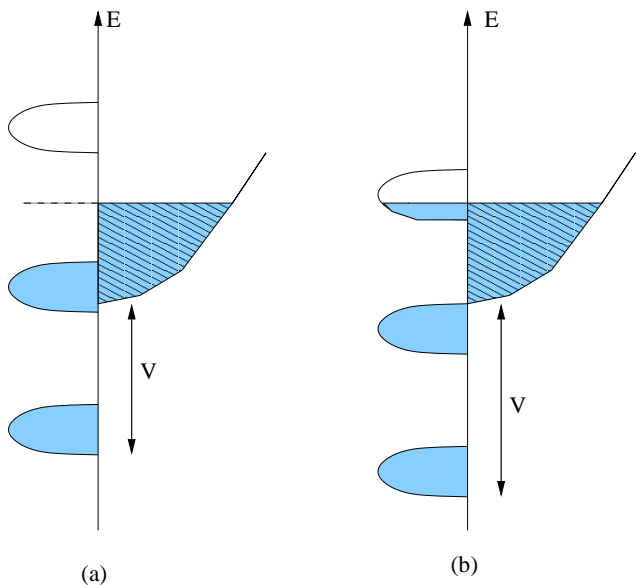}
\end{center}
\caption{ (Color online) Schematics of the occupation of atoms and 
molecules near 
a Feshbach 
resonance when $U_b$ is large.
(a) is when $X$ is not inside the window of $[0,1]$; (b) is for $X \in [0,1]$.  
($X$ is defined in Eq.(3).)
In both (a) and (b), fully shaded lobes stand for completely filled 
molecular Mott bands and empty lobes for unfilled Mott bands; 
the partially filled lobe is for a partially filled band.
The patterned area to the right of the vertical axis in (a) or (b) is 
the partially filled atomic band.
The distance from the bottom of the atomic band and the bottom of the lowest 
molecule Mott band
is the detuning $|v|$.
\label{Fig1}}
\end{figure}

A finite but small value of $t_b$ won't change the general 
structure of the step-like function, except that the 
transition from the $I$th plateau to the $(I+1)$th plateau becomes smooth.
Generally, we have

\begin{eqnarray}
&& N_b(2\mu-2v; t_b, U_b)=
\nonumber \\
&&Int \frac{2\mu-2v}{U_b}
+F\left(\frac{2\mu-2v -Int(2\mu-v/U_b) 
U_b}{2zt_b}+\frac{1}{2}\right),\nonumber 
\\
&& F(X)=\left\{\begin{array}{cc}
1, & \mbox{$1\leq X < \frac{U_b}{4zt_b}$};\\
f(X), & \mbox{$ 0\leq X \leq 1$}; \\
0, & \mbox{$ -\frac{U_b}{4zt_b} \leq X \leq 0$}.
\end{array}\right.
\end{eqnarray}
Here $Int Y=I$ if $ -1/2 \leq  Y-I  < 1/2 $; 
$X$ is the distance (in energy space as shown in Fig.\ref{Fig1}) 
from the 
chemical potential to the bottom of the {\em nearest} Mott band measured 
in the unit of $2zt_b$ (where $z$ is the coordination number). The filling 
factor $f(X)$ is a continuous function 
of $X$ varying from $0$ at $X=0$ to one at $X=1$.
Its detailed form depends on the band structure, 
which is not specified here.

\begin{figure}[tbp]
\begin{center}
\includegraphics[width=3in]
{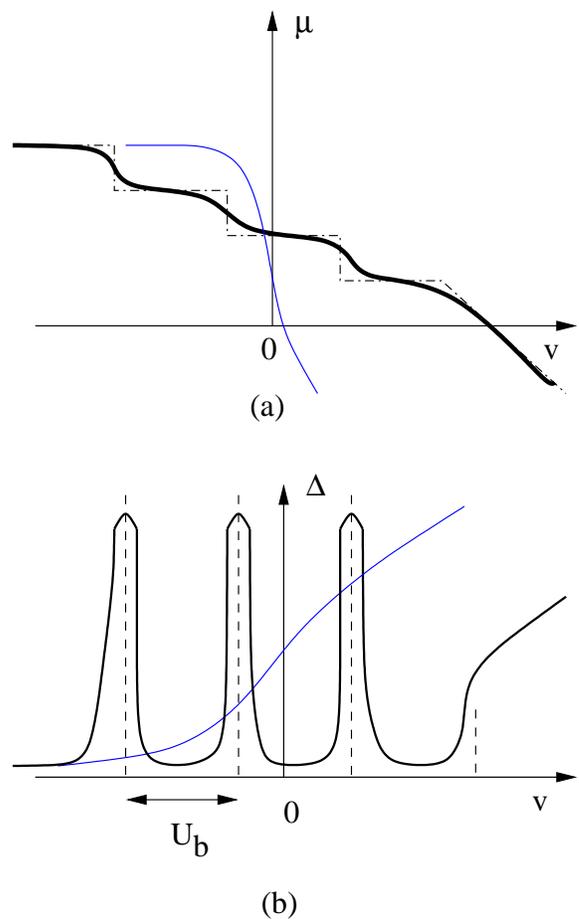}
\end{center}
\caption{ (Color online) Schematics of solutions obtained in this letter 
for $N/2=3+\epsilon$
($0<\epsilon<1$).
(a) the $v$-dependence of the chemical potential $\mu$ for atoms.
The thick line is for the limit when $U_b$ is much larger than $t_b$;
the dashed-dotted line represents the limit when $t_b$ approaches zero.
The thin blue line is the typical chemical potential in the weakly interacting limit when 
$U_b$ is much smaller than $t_b$.
(b) The thick curves show the $v$-dependence of the BCS quasi-particle gap
when $U_b$ is large. The gap saturates at the value given after 
Eq.(\ref{gap2}) 
when approaching the dashed lines.
Dashlines locate the position of the center of a Mott band along the 
$v$-axis.
In both (a) and (b), the thin blue curves are for the limit when $U_b$ 
vanishes. 
\label{Fig2}}
\end{figure}

Taking into account this general property of $N_b$, 
one easily finds that the solution to
Eq.(\ref{cpotential}) is also a step-like function of $v$ (see 
Fig.\ref{Fig2}).
Whenever the detuning is such that 
$-\frac{U_b}{4zt_b} < X \leq 0$ or $1\leq X <\frac{U_b}{4zt_b}$, 
varying the detuning $v$ does not change $N_b$ or $N_f$ and 
therefore the chemical potential for fermions must be a 
constant\cite{remark1}. This 
leads to plateaus in the 
self-consistent chemical potential. 
The first plateau represents the limit of extremely weakly interacting
superfluids, i.e. the detuning is positive and much larger
than the fermion energy of atoms. The chemical 
potential in this limit barely depends on the 
detuning $v$ or coupling constant $\gamma_0$;
its value is approximately equal 
to the Fermi energy of a noninteracting Fermi gas, i.e. $\mu_1\approx 
\epsilon_F$; $\epsilon_F$ is the Fermi energy
of a gas with N atoms per site.
The value of the chemical potential on the $i$th plateau is given as

\begin{eqnarray}
\mu^{3/2}_i =\mu^{3/2}_1 - (i-1) (\epsilon_F^0)^{3/2}, i \leq i_{max}
\end{eqnarray}
where $\epsilon_F^0$ is the Fermi energy of an atomic gas with  
two atoms per lattice site.
The number of plateaus in the chemical potential $i_{max}$ is determined by the density of
atoms; for $N/2=n+\epsilon$ ($n$ is an integer and $0< \epsilon <1$), 
the value of $i_{max}$ and the number of valleys 
in Fig.\ref{Fig2}(b) are $n+1$.   
 
The transition between two plateaus takes place when 
the self-consistent chemical potential is located inside a Mott 
band; i.e. when $X$ is between zero and one. 
The width of the transition region
when the detuning varies is proportional to the band width $2zt_b$.
The distance between two consecutive transition regions is
$v_{i+1} - v_{i}=U_b$;
$v_{i}$ is the detuning at the $i$th transition region when 
$X(v_i)=1/2$.

The trial wavefunction for the fermion sector is 

\begin{eqnarray}
&& |g.s.> = \prod_{\bf q} [ u({\bf q}) +v_{\eta\xi}({\bf q}) 
f^{\dagger}_{\eta\uparrow}({\bf q}) f^\dagger_{\xi\downarrow}(-{\bf q}) 
]|0>;
\end{eqnarray}
${\bf q}$ is the crystal quasi-momentum
and $u^2+\sum_{\eta\xi} v^2_{\eta\xi}=1$.
The solutions when chemical potentials are located on a plateau are 
distinctly different from the solutions when the chemical potentials are 
in a smooth transition region (Fig.\ref{Fig2}(a)). We will discuss these 
two 
limits separately.
When $X$ (defined in Eq.(3)) is either larger
than unity or smaller than zero,
the chemical potential falls on a plateau.
Unoccupied molecule states, with which fermions in extended states have 
resonances,
are well above the self-consistent 
chemical potential.
The difference between the energy of the lowest molecule state and 
the chemical potential
depends on the detuning $v$ or $X$. 
The energy difference $2 \delta_1(X)z t_b (>0)$ is equal to $- 2X zt_b$ 
if $ 0> 
X > -U_b/4zt_b$
and equal to $U_b - 2X zt_b$ if $  1< X < U_b/4zt_b$;
typically it is a fraction of 
$U_b$.
When $\delta_2(X) \gg 1$, one obtains the effective on-site 
attractive interaction between atoms mediated by molecules 

\begin{eqnarray}
V_{\eta\xi,\eta'\xi'}=-{\lambda_f} g_{\eta\xi} 
g_{\eta'\xi'}, 
\lambda_f=\frac{\gamma_0^2}{ 2 \delta_1(X) z t_b}.
\end{eqnarray}
Minimization of the free energy $< {\cal H} > - \mu N$ leads to the 
following equations for
$(u, v_{\eta\xi})$($V$ is the volume):

\begin{eqnarray}
&& 2 u v_{\xi\eta'} 
\xi_{\eta'\eta}({\bf q})
=\Delta (g_{\eta\xi} u^2 - g_{\eta'\xi'}v_{\eta'\xi'} 
v_{\eta\xi});\nonumber \\
&& \Delta =\frac{{\lambda}_f}{V} 
\sum_{{\bf q},\eta,\xi} g_{\eta\xi}v_{\eta\xi}({\bf q}) 
u({\bf q}).
\end{eqnarray}
Here 
$\xi_{\eta\eta'}({\bf q})=\epsilon_{\eta\eta'}({\bf q})-\mu$,
$\epsilon_{\eta\eta'}$ is the kinetic energy tensor. 
The lattice constant has been set to unity.

Solutions to this equation depend on the details of 
$g_{\eta\xi}$ and the dispersion
$\xi_{\eta\eta'}({\bf q})$. In the limit when the different bands 
are degenerate, i.e. $\xi_{\eta\eta'}({\bf q})=\delta_{\eta\eta'}\xi({\bf q})$, one 
obtains $v_{\eta\xi}=g_{\eta\xi}v$;
$(u, v)$ are the usual BCS solutions,

\begin{eqnarray}
&& \frac{2 uv}{u^2-v^2}=\frac{\Delta}{\xi({\bf q})}; \nonumber\\
&& \frac{V}{{\lambda}_f}=\sum_{\bf q} \frac{1}{2E({\bf q})}-\sum_{\bf q} 
\frac{1}{2\epsilon({\bf q})}.
\end{eqnarray}
Here $E({\bf q})=\sqrt{\xi^2({\bf q})+|\Delta|^2}$ is the quasi-particle 
energy;
$\sum_{\eta\eta'}g^2_{\eta\eta'}=1$. 
The energy gap
$\Delta \sim t_f \exp(-2/\tilde{\lambda}_f)$ is exponentially 
small. 
$\tilde{\lambda}_f$ is a dimensionless interaction constant 
estimated 
to be $C \frac{\gamma_0^2 N_f^{1/3} 
}{2\delta_1(X)z t_b t_f}$;
we have assumed the atomic density of states at the chemical potential is
$C N_f^{1/3}/t_f$, $C$ is a quantity of order unity which depends on the 
details of the band structure.

When $0\leq X \leq 1$,
the chemical potential is between two plateaus and varies continuously as 
a function of $X$ within this window. In
the mean field approximation one obtains the following 
modified self-consistent equation,
$\Delta=\gamma_0 <b>$, 
$- \delta_2(X) 2zt_b + \lambda <b>^2=$
$\gamma_0^2 \sum_{\bf q} ( {1}/{2E({\bf q})}-{1}/{2\epsilon({\bf q})} )$;
$\lambda (U_b)$ is an interaction constant appearing in an 
effective theory and is of order of $U_b$
\cite{Zhou05b}; $\delta_2(X)$ is equal to $X$ when $ 0 \leq X\leq 1$.
In the leading order, one neglects the 
$\gamma_0$-dependence of $<b>$ and $\Delta$ is linearly 
proportional to $\gamma_0$. So 
in the dilute limit ($X \ll 1$), $<b>$ is of order of 
$\sqrt{\delta_2(X) 2 z t_b / U_b}$. 
However 
once the chemical potential of interacting molecules
$\lambda(U_b) N_b$  becomes comparable to
the Fermi energy of a Fermi gas with the corresponding density
(which scales as $N_b^{2/3} t_b$), the mean field approach fails;
this sets the limit on
the number of atoms per lattice site, which has to be much less than 
$(t_b/U_b)^3$.
The window where mean field theory is applicable thus shrinks to zero
as $t_b/U_b$ approaches zero.

To get an estimate of $<b>$ in the large-$U_b$ limit, 
it is useful to map the model to an 
XXZ model. 
Since the completely filled Mott bands are irrelevant to the discussions 
of superfluids, we only need to take into account the partially filled 
band. 
When $t_b$ is much less than $U_b$, the molecules can be treated as 
{\em hard-core bosons} at energy scales much lower than $U_b$.
For hard-core bosons, the on-site Hilbert space is
one boson and no bosons. The average number of bosons at each lattice site 
varies from zero to one depending on the position of chemical 
potentials. This suggests an effective pseudo spin $S=1/2$ 
subspace (on-site): 
$ |\sigma^z_{k}=-1>=b^\dagger_k |vac>_b$,
$ |\sigma^z_{k}=1>=|vac>_b$.
$|vac>_{b}$ is the vacuum of bosons.
Furthermore, in this truncated space one can easily verify that
three Pauli matrices can be represented as 
$\sigma^z_k \rightarrow 1-2 b^\dagger_k b_k$, $\sigma^{+}_k 
\rightarrow b_k$
and $\sigma^{-}\rightarrow b_k^\dagger $;
$\sigma^{\pm}=({\sigma^x\pm i\sigma_y})/2$\cite{Zhou05a,Zhou05b}.

A state with $\sigma^z_k=1 $ corresponds to a vacuum of molecules,
and $\sigma^z_k=-1$ a
Mott state with
one molecule per lattice site.
On the other hand, in translationally invariant
states where $\sigma_{k}$ has a finite expectation value in the {\em XY} plane, 
the condensation amplitude of molecules is nonzero; i.e.

\begin{eqnarray}
&& < b_k >=<{\sigma}^-_k>.
\end{eqnarray}
So any state with all $<\sigma_k>$ pointing along a specified 
direction that differs from the $z$-axis breaks the in-plane (XY 
type) 
rotational symmetry 
of the pseduo spin and corresponds to a superfluid. 
The effective Hamiltonian for molecules can then be written as

\begin{eqnarray}
&& H_{eff}=
-\frac{t_b}{2} \sum_{<kl>} {\sigma}^x_{k} {\sigma}^x_{l} + 
{\sigma}^y_{k} {\sigma}^y_{l}
+ h^z_b \sum_k \sigma^z_{k}.
\end{eqnarray}
Here $h^z_b=(X-1/2) 2 z t_b$.
The Hamiltonian is invariant under a rotation around the $z$-axis
or has an $XY$ symmetry which corresponds to the usual $U(1)$ symmetry
of a boson field.

One easily finds the ground state of the above Hamiltonian using 
a semiclassical approximation\cite{Zhou04}
and
gains insight into superfluids in the large-$U_b$ limit.
In the semiclassical approximation, 
one assumes that the orientation of a pseudo spin is specified by a polar
angle $\Theta$ and an azimuthal angle $\Phi$ as: 
$<\sigma^x>=\sin\Theta \cos\Phi$,
$<\sigma^y>=\sin\Theta\sin\Phi$ and $<\sigma^z>=\cos\Theta$.

Minimizing the mean field energy with respect to $\Theta$ leads to
the following solution for $\Theta$.
$\Theta$ remains at $0$ as $X$ becomes negative,
corresponding to an empty Mott band.
As $X$ becomes larger than 1, $\Theta$ is equal to $\pi$ 
representing a Mott state with one molecule per site.
For $1> X>0$, one obtains $\cos\Theta=1-2X$, or 
$\sin\Theta=2 \sqrt{X(1-X)}$; 
note that in this limit all solutions with a given $\Theta$ but arbitrary 
$\Phi$ 
are degenerate and break the usual $U(1)$
symmetry of a superfluid. 
Following Eq.(9) and Eq.(10), one arrives at
$\Delta=\gamma_0 
\sin {\Theta}\exp(- i\Phi)$, or in terms of $X$

\begin{eqnarray}
\Delta= \gamma_0 2\sqrt{X(1-X)} \exp(- i\Phi).
\label{gap2}
\end{eqnarray}
At $X=1/2$ or half filling,  
$\sin\Theta=1$ and the modulus of the expectation value 
$< b >$ (proportional to $\sin\Theta$) 
reaches its maximum. 
This leads to the maximal value of the BCS gap,
$|\Delta| = \gamma_0$.

The calculation of $T_c$ in the large $U_b$ limit is tricky because of
large quantum fluctuations present when the chemical potential falls in 
the middle 
of a Mott band.
Here we estimate $T_c$ again in two limits: a) when the chemical 
potential falls on a plateau corresponding to various valleys in 
Fig.\ref{Fig2} 
(b);
b) when the chemical potential is between two plateaus.
In case a), fermions are weakly interacting and we expect the BCS theory 
is valid. $T_c$ scales as $\Delta$, the solution to Eq.(8) and is 
exponentially small when $\gamma_0$ is much smaller compared to $U_b$. In 
case b), $T_c$ 
should be 
approximately the BEC transition temperature of molecules in a 
partially filled Mott band, which is proportional to $\epsilon^{2/3}t_f$
($\epsilon (<1)$ is the filling fraction.).  This temperature
differs from the quasi-particle gap calculated in Eq.(\ref{gap2}) and is 
also independent of the coupling constant $\gamma_0$.
It reaches a maximal value when $X=1/2$ because of the 
particle-hole symmetry.
Therefore, we expect an oscillatory behavior in the critical temperature
as well. At a temperature lower than the BEC temperature but higher than 
the exponentially small BCS transition temperature when the chemical 
potential is between two 
Mott bands, the ground state is expected to alternate between a 
fermionic superfluid 
and a normal fluid as the detuning is varied. This might be observed in 
future experiments on {\em LFR}s. 
I would like to thank T. Leggett, K. Madison and G. Warner for  
discussions. This work is supported by a grant 
from UBC, a Discovery grant from NSERC,
Canada and by the A. P. Sloan foundation.


\begin{thebibliography}{99}
\bibitem{remark2}Strictly speaking, molecular Mott states are not 
well-defined in the presence of resonances between fermion atoms and 
bosonic molecules. See
more discussions in Ref.\cite{Zhou05a,Zhou05b}. However, in the current situation, we find the induced  
condensation fraction in a molecular Mott state
is of order of $(\gamma_0/U_b) \exp(-2/\tilde{\lambda_f})$ and for practical reasons we neglect 
such a contribution in our calculations because $U_b$ is assumed to be
much larger than $\gamma_0$.
\bibitem{remark1}In this limit, the interactions between atoms 
mediated by molecules are negligible so that 
at a given chemical potential, we can neglect 
the $v$-dependence in the calculation of $N_f$.
\end{thebibliography}
\end{document}